\documentclass[10pt,twocolumn,article]{IEEEtran}
\usepackage{etex}
\usepackage{tikz}

\usetikzlibrary{shapes,arrows,trees,calc}

\usetikzlibrary{arrows,shapes,positioning,shadows,trees}

\tikzset{
  basic/.style  = {draw, text width=2cm, drop shadow, font=\sffamily, rectangle,},
  root/.style   = {basic, rounded corners=2pt, thin, align=center,
                   fill=green!30},
  level 2/.style = {basic, rounded corners=6pt, thin,align=center, fill=green!60,
                   text width=12em},
  level 3/.style = {basic, thin, align=left, fill=pink!60, text width=8.5em}
}

\usepackage{amsfonts}
\usepackage{stfloats}   
\usepackage[T1]{fontenc}
\usepackage{setspace}
\usepackage{cite}
\usepackage[latin1]{inputenc}
\usepackage{epsf}\epsfverbosetrue
\usepackage{graphics,epsfig,color}
\usepackage{graphicx,epsfig}
\usepackage{epsfig}
\usepackage{multirow}
\usepackage{slashbox}
\usepackage{alltt}
\usepackage{subfigure}
\usepackage{color}
 \usepackage{float} 
\usepackage{url}
\usepackage{graphicx}
\usepackage{verbatim} 
\usepackage{latexsym} 
\usepackage{amsmath}
\usepackage{color}
\usepackage{wrapfig}
\usepackage{algorithm}
\usepackage{eso-pic}
\usepackage{fix-cm}

\usepackage{color}
\usepackage{wrapfig}
\usepackage{algorithm}
\usepackage{layout}
\usepackage{amsmath}
\usepackage{textcomp}
\usepackage{multirow}
\usepackage{soul,xcolor}
\usepackage{amsmath}
\usepackage{verbatim}

\usepackage{tikz}
\usetikzlibrary{shapes}
\usepackage{amsmath}
\usepackage{xspace}

\renewcommand{\theenumi}{\alph{enumi}}  

\usepackage{tikz}
\makeatletter
\@namedef{color@1}{red!40}
\@namedef{color@2}{green!40}
\@namedef{color@3}{blue!40}
\@namedef{color@4}{cyan!40}
\@namedef{color@5}{magenta!40}
\@namedef{color@6}{yellow!40}

\newcommand{\graphitemize}[2]{%
\begin{tikzpicture}[every node/.style={align=center}]
  \node[minimum size=5cm,circle,fill=gray!40,font=\Large,outer sep=1cm,inner sep=.5cm](ce){#1};
\foreach \gritem [count=\xi] in {#2}
{\global\let\maxgritem\xi}
\foreach \gritem [count=\xi] in {#2}
{%
\pgfmathtruncatemacro{\angle}{360/\maxgritem*\xi}
\edef\col{\@nameuse{color@\xi}}
\node[circle,
     ultra thick,
     draw=white,
     fill opacity=.5,
     fill=\col,
     minimum size=3cm] at (ce.\angle) {\gritem };}%
\end{tikzpicture}
}%

\newcommand{\blue}[1]{\textcolor{black}{#1}}

\newcommand*{\rom}[1]{\expandafter\@slowromancap\romannumeral #1@}
\newcommand{\Rmnum}[1]{\expandafter\@slowromancap\romannumeral #1@}

\begin{document}
\setstcolor{red}
\title{NOMA versus OMA in Finite Blocklength Regime: Link-Layer Rate Performance}
\author{Muhammad Amjad,~\IEEEmembership{Student Member,~IEEE,} Leila Musavian,~\IEEEmembership{Member,~IEEE, }and Sonia A\"{i}ssa,~\IEEEmembership{Fellow,~IEEE}
\thanks{M. Amjad and L. Musavian are with the School of Computer Science and Electronic Engineering, University of Essex, CO4 3SQ, UK (Email: \{m.amjad, leila.musavian\}@essex.ac.uk). S. A\"{i}ssa is with the Institut National de la Recherche Scientifique (INRS-EMT), University of Quebec, Montreal, QC, H5A 1K6, Canada (Email: aissa@emt.inrs.ca).}}
\maketitle

 \thispagestyle{empty}

\begin{abstract}
In this paper, we investigate the latency performance of non-orthogonal multiple access (NOMA) and orthogonal multiple access (OMA) technologies in finite blocklength regime. In the comparative study, we derive the achievable effective capacity of two-user NOMA and its OMA counterpart under delay quality-of-service constraints. We then obtain closed-form expressions for the achievable effective capacity of the weak and strong users in both scenarios considering transmissions over Rayleigh fading channels. Numerical results are provided. In particular, it is shown that at low signal-to-noise ratios (SNRs), the OMA user with better channel condition outperforms both NOMA users. \blue{We also evaluate the impact of fixed power allocation scheme on the achievable effective capacity of two-user NOMA.} The comparative analysis of the total link-layer rate shows that at high SNRs, the total link-layer rate of NOMA with finite blocklength outperforms the one of OMA when the delay exponent is loose.
\end{abstract}
\section{Introduction}
\label{sec:in}
Transition from ultra-low latency to massive ultra reliable and low latency communications (mURLLC) for beyond 5G (B5G) applications demands the researchers from both industry and academia to revisit the enabling technologies. Future smart cities, autonomous robotics, holographic communications, blockchain, and massive sensing are few examples to name that require the mURLLC service class of B5G \cite{R173}. However, achieving mURLLC for the target future applications is a challenging task. While the non-orthogonal multiple access (NOMA) in conjunction with finite blocklength (short packet) communications is considered as an enabler for low-latency communications \cite{R160}, further research is required to quantify the end-to-end latency in these systems. Also, the scalability of this technology is yet to be investigated.

NOMA with finite blocklength has the potential to allow ultra-low latency, massive connectivity, and higher throughput under favorable conditions \cite{R166}. The principle of NOMA in finite blocklength regime follows the conventional concept of NOMA, with superposition coding at the transmitter and successive interference cancellation (SIC) at the receiver \cite{R177}. However, when operating with finite blocklength packets, the Shannon formula is not a good approximate for the achievable rate of NOMA, and alternative solutions are needed. In this vein, the authors in \cite{R139} provided a framework to approximate the achievable rate of a point-to-point communication link in finite blocklength regime.

Can the latency requirements of mURLLC for B5G services be satisfied with NOMA in finite blocklength regime? This question needs a detailed delay performance analysis of NOMA in finite blocklength regime. In this regard, the authors in \cite{R140} investigated the performance of NOMA with short-packet communications subject to reliability constraint. More specifically, the mentioned work showed the reduction in physical-layer transmission latency while using NOMA in conjunction with short-packet communications. The latency performance of NOMA with finite blocklength was further investigated in \cite{R176}, which confirmed the improved performance of NOMA in terms of reducing latency and improving throughput, in comparison to orthogonal multiple access (OMA). 

Also, a comparative view of the achievable effective capacity (EC) of uplink two-user NOMA and OMA was conducted in \cite{R179}, but not for transmissions in finite blocklength regime. Later in \cite{R156}, the achievable EC for systems with finite blocklength was analyzed, and it was shown that the proposed system within short blocklength and reliability constraint can reduce latency, hence establishing the importance of short-packet communications for achieving low latency. Focusing on the importance of short packet communications, the achievable EC for finite blocklength machine-type communications (MTC) under delay constraint was derived in \cite{R178}. In that work, the optimum error probability was characterized under the effect of signal-to-noise ratio (SNR) variations to maximize the achievable EC, and it was confirmed that under strict delay constraints, the SINR variations have less effect on the achievable EC of MTC.

In this work, we investigate the latency performance of NOMA and OMA with short-packet communications. The major contributions of this paper can be summarized as follows: {\it (i)} We derive the achievable EC of two-user NOMA and OMA in finite blocklength regime.\footnote{Two-user NOMA has been included as a building block in third generation partnership project long-term evolution advanced (3GPP-LTE-A) networks \cite{R175}.} Specifically, the achievable EC (link-layer rate) of NOMA users is investigated in finite blocklength regime under heterogeneous delay quality-of-service (QoS) requirements, in comparison with the OMA counterpart. {\it (ii)} We further obtain closed-form expressions for the individual users' EC in the two-user NOMA and OMA networks, and confirm their accuracy using Monte-Carlo simulations. {\it (iii)} We show that the OMA user with better channel conditions outperform both NOMA users at low SNRs, while the total link-layer rate of NOMA outperforms the one of OMA under loose delay constraints.
\section{Transmission Framework and Fundamentals}
\label{sec:system_model}
Consider a downlink two-user NOMA network with finite blocklength. The users, denoted by $v_i$, $i=\{1,2\}$, are equipped with single antennas and communicate with a single base station (BS). The channel coefficient between the BS and $v_i$ at time $\tau$ is referred to by $h_i(\tau)$. The two users are classified based on their channel conditions as strong and weak users and, without loss of generality, we assume $\left | h_1 (\tau)  \right |^2 \geq \left | h_2 (\tau) \right |^2$. 

Following the NOMA principle, the BS broadcasts a combined message $\sum_{i=1}^{2}\sqrt{\alpha_{i}P}u_{i}(\tau)$ to its users, where $u_i$ is the message corresponding to user $v_i$, $P$ is the BS's total transmit power, and $\alpha_{i}$ is the power coefficient for user $v_i$. \blue{With fixed power allocation policy at the BS,} the power coefficients for the two users are such that $\alpha_{1}\leq\alpha_{2}$. The received signal at user $v_{i}$ can now be formulated as \blue{$y_i=h_i\sum_{i=1}^{2} \sqrt{\alpha_{i}P}u_i + m_i$},\footnote{As the channel coefficients are assumed stationary and ergodic random processes, the time index $\tau$ is omitted hereafter for simplicity of presentation.} where $m_i$ is the additive white Gaussian noise (AWGN) at $v_i$, $i\in\{1,2\}$.

At the receiving side, the strong user ($v_1$) first performs SIC to remove interference ($u_2$) from its received signal $(y_1)$, and then decodes its own message. Therefore, for user $v_1$, the received SNR, denoted by ${\rm{SNR}_{1}^{N}}$,\footnote{Superscript N indicates NOMA. Later, notation O will be used to indicate the OMA operation.} can be found as
\begin{equation}
\label{eq:snr_u}
 {\rm{SNR}_{1}^{N}}=\alpha_{1}\rho\left | h_1 \right |^2 ,
\end{equation}
where $\rho$ is the transmit SNR, namely $\rho=\frac{P}{N_{o}B}$, in which ${N_{o}B}$ denotes the noise power. 

On the other hand, the weak user $(v_2)$ treats $u_1$ as interference and decodes its own message directly. Hence, its resulting signal-to-interference-plus-noise ratio (SINR) can be derived as
\begin{equation}
\label{eq:sinr_t}
{\rm{SINR}_{2}^{N}}=\frac{\alpha_{2} \rho \left | h_{2} \right |^2 }{\alpha_{1} \rho \left | h_{2} \right |^2+1}.
\end{equation}

Channel gains of both users are modeled as Rayleigh distributions with unit variance. Following the NOMA operation, the users $v_1$ and $v_2$ are sorted based on their ordered channel gains. Therefore, the probability density function (PDF) of the ordered channel power gains can be obtained using the order statistics \cite{R163}. In this regard, using $\rho \left | h_i \right |^2=\gamma_i$ and denoting its PDF as $f\left ( \gamma_{i} \right )$, we apply the order statistics to get
\begin{equation}
\begin{aligned}
\label{eq:pdf_sw}
f_{\gamma_{1:2}}\left ( \gamma_{1} \right )&=\xi_{1}f\left ( \gamma_{1} \right )F\left ( \gamma_{1} \right ),\\
f_{\gamma_{2:2}}\left ( \gamma_{2} \right )&=\xi_{2}f\left ( \gamma_{2} \right )\left (1-F\left ( \gamma_{2} \right )       \right ),
\end{aligned}
\end{equation}
\noindent where $f_{\gamma_{i:2}}$ is the PDF of the ordered $\gamma_i$ out of two users,   $\xi_{i}=\frac{1}{B\left ( i,2-i+1 \right )}$, in which  $B(a,b)$ is the Beta function \cite{R171}, and $i\in\{1,2\}$.

For the case with OMA operation, both users have access to the same spectrum bandwidth as in the NOMA case but each user can only occupy half of the transmission time slot.  Using the results of \cite{R139} as starting point, the users' achievable rates with finite blocklength in the NOMA and OMA cases under study can be formulated, in b/s/Hz, as
\begin{equation}
\label{eq:rate_nomas}
r_{1}^{{\rm{N}}}= {\rm log_2}\left(1+\alpha_{1}\gamma_1 \right)-\sqrt{\frac{V_{1}^{{\rm{N}}}}{n}}Q^{-1}{(\epsilon)},\ \  \qquad\qquad\qquad
\end{equation}
\begin{equation}
\label{eq:rate_nomaw}
r_{2}^{{\rm{N}}}={\rm log_2}\left( 1+\frac{\alpha_{2}\gamma_2 }{\alpha_{1}\gamma_2+1} \right)  -  \sqrt{\frac{V_{2}^{{\rm{N}}}}{n}}Q^{-1}{(\epsilon)},\qquad\qquad
\end{equation}
\begin{equation}
\label{eq:rate_omas}
r_{i}^{{\rm{O}}}= \frac{1}{2} \left( {\rm log_2}\left ( 1+ \gamma_i \right )  -  \sqrt{\frac{V_{i}^{{\rm{O}}}}{n}}Q^{-1}{(\epsilon)} \right ), \ i\in\{1,2\},
\end{equation}
where $r_{1}^{{\rm{N}}}$, $r_{2}^{{\rm{N}}}$ and $r_{i}^{{\rm{O}}}$ are the achievable rates of the NOMA strong user, NOMA weak user, and OMA users, respectively, $n$ is the blocklength, $\epsilon$ is the transmission error probability, and  $Q^{-1}(.)$ is the inverse of Gaussian Q-function with $Q\left ( x \right )=\int_{x}^{\infty}\frac{1}{\sqrt{2\pi}}e^{-\frac{w^2}{2}}dw$ \cite{R172}. Also, \blue{$V_{1}^{{\rm{N}}}=1-\left ( 1+\alpha_{1}\gamma_1 \right )^{-2}$},  \blue{$V_{2}^{{\rm{N}}}=1-\left(1+\frac{\alpha_{2}\gamma_2 }{\alpha_{1}\gamma_2+1} \right)^{-2}$}, and \blue{$V_{i}^{{\rm{O}}}=1-\left ( 1+\gamma_i \right )^{-2}$} are the channel dispersions of the NOMA strong user, NOMA weak user, and OMA users, respectively.
\subsection{Theory of Effective Capacity}
In this subsection, we explain the basic concepts related to the theory of EC. This metric is used to find the maximum arrival rate for a given service rate while satisfying a certain delay-outage probability constraint \cite{R158}. 

We assume that the transmission scheme in our network is required to satisfy statistical delay QoS constraints. It is shown that if a queue length exceeds a certain threshold $(x)$, then by using the large deviation theorem \cite{R167} the probability of buffer overflow will hold the following equality
\begin{equation}
\label{eq:ec_theta1}
-\lim\limits_{x \to \infty}\frac{{\rm{ln}}\left ({\rm{Pr}}\left \{q_i(\infty)> x  \right \}  \right )}{x}=\theta_i,
\end{equation}
where $q_i(\infty)$ is the steady-state transmit buffer of user $v_i$, $\theta_i$ is this user's delay exponent, and ${\rm{Pr}\{a\textgreater b\}}$ is the probability that $a \textgreater b$ holds.  Following (\ref{eq:ec_theta1}), the queueing delay violation probability can be estimated as \cite{R158}
\begin{equation}
 \label{eq:qdvp}
 {\rm{Pr}}\left \{  D_i>D_{\rm{max}}^i\right \}\approx {\rm{Pr}}\left \{  q_i(\infty)>0\right \}e^{-\theta_i\mu_i D_{\rm{max}}^i},~i\in\{1,2\},
\end{equation}
where, for user $v_i$, $D_{\rm{max}}^i$ is the maximum delay, ${\rm{Pr}}\left \{ q_i(\infty)>0 \right \}$ is the probability of non-empty buffer, and $\mu_i$ is the maximum arrival rate.
\section{Effective Capacity of NOMA and OMA in  Finite Blocklength Regime}
\label{sec:ec_noma_oma}
In this section, we derive the achievable EC of the two-user NOMA and OMA networks described above in finite blocklength communication regime. We then provide closed-form expressions for the EC.

By following  \cite{R156} and \cite{R154}, the achievable EC of the two-user NOMA and the OMA counterpart in finite blocklength regime can be formulated as
\begin{equation}
\label{eq:1EC}
C_{i}^{{\rm{N}}}= -\frac{1}{\theta_{i}{n}}{\rm{ln}} \left( \mathbb{E} \left[\epsilon+\left ( 1-\epsilon \right )e^{-\theta_{i}{n}{r_{i}^{{\rm{N}}}}}  \right] \right) ,
\end{equation}
\begin{equation}
\label{eq:1EC}
C_{i}^{{\rm{O}}}= -\frac{1}{\theta_{i}{n}}{\rm{ln}} \left(\mathbb{E} \left[\epsilon+\left ( 1-\epsilon \right )e^{-\theta_{i}{n}{r_{i}^{{\rm{O}}}}}  \right] \right),
\end{equation}
\noindent where $C_{i}^{{\rm{N}}}$ and $C_{i}^{{\rm{O}}}$ represent the EC of user $v_i$ in finite blocklength regime, for NOMA and OMA, respectively, and $\mathbb{E}[.]$ is the expectation operator.

By considering the service rate ${r_{i}^{{\rm{N}}}}$ for users $v_i$ in finite blocklength regime from (\ref{eq:rate_nomas}) and (\ref{eq:rate_nomaw}), the achievable EC of the NOMA strong user and the NOMA weak user can be approximated as
\begin{align}
 \label{eq:ec_nomas}
C_{1}^{{\rm{N}}}= -\frac{1}{\theta_{1}{n}}{\rm{ln}} \left(\mathbb{E} \left[\epsilon+\left ( 1-\epsilon \right )
\left ( 1+\alpha_{1}\gamma_1 \right )^{2\Upsilon_{1}} e^{\psi_{1}{\sqrt{V_{1}^N}}}  \right] \right),
\end{align}
\noindent and
\begin{align}
 \label{eq:ec_nomaw}
C_{2}^{{\rm{N}}} = &-\frac{1}{\theta_{2}{n}} \nonumber\\
&\times {\rm{ln}}\left(   \mathbb{E} \left [\epsilon+\left ( 1-\epsilon \right ) \Big(1+   \frac{\alpha_{2}\gamma_2 }{\alpha_{1}\gamma_2+1}  \Big)^{2\Upsilon_{2}}   e^{\psi_{2}{\sqrt{V_{2}^{N}}}}  \right ] \right) ,
\end{align}
\noindent respectively, where $\Upsilon_{i}=-\frac{\theta_{i}{n}}{2{\rm{ln}2}}$, and  $\psi_{i}=\theta_i\sqrt{n}Q^{-1}{(\epsilon)}$.

As specified, users $v_1$ and $v_2$ can also operate according to OMA, by transmitting their messages using time division multiple access (TDMA). For the OMA case, using (\ref{eq:rate_omas}) the achievable EC of the two users can be approximated as
\begin{align}
\label{eq:ec_omas}
{C}_i^{{\rm{O}}}=-\frac{1}{\theta_{i}{n}}{\rm{ln}} \left( \mathbb{E} \left[\epsilon+\left(1-\epsilon \right)
\left(1+\gamma_i \right)^{\Upsilon_{i}} e^{\frac{\psi_{i}{\sqrt{V_{i}^O}}}{2}}\right] \right).
\end{align}

The above derived individual EC expressions of the two users with NOMA or OMA in finite blocklength regime can be used to analyze and compare the delay performance in both operation scenarios.

To further simplify the above expressions, we derive closed-form expressions for the individual EC of the strong and weak NOMA and OMA users in finite blocklength regime. Specifically, using the order statistics from (\ref{eq:pdf_sw}), the final closed-form expressions for the two users, in NOMA and OMA, can be obtained as shown in (\ref{eq:ec_noma_s_closed}) to (\ref{eq:ec_oma_w_closed}), \blue{where ${\rm{H}}\left ( a,b,z \right )=\frac{1}{\Gamma\left ( a \right )} \int_{0}^{\infty}e^{-zt}t^{a-1}\left ( 1+t \right )^{b-a-1} dt$ is the confluent hypergeometric function of the second kind \cite{R171}, and ${\rm{E_i}}(.)$ is the exponential integral \cite{R172}.} Details on the derivation of the closed-form expressions for $C_1^{{\rm{N}}}$, ${C}_1^{{\rm{O}}}$ and ${C}_2^{{\rm{O}}}$ are provided in Appendix A, while the proof for deriving the closed-form expression for $C_2^{{\rm{N}}}$ is presented in Appendix B.
\begin{figure*}
\begin{align}
&\begin{aligned}
\begin{split}
 \label{eq:ec_noma_s_closed}
C_{1}^{{\rm{N}}}= -\frac{1}{\theta_{1}{n}}{\rm{ln}}\Bigg ( \epsilon+\left ( 1-\epsilon \right )   \frac{2}{\alpha_{1} \rho }e^{\psi_{1}}\left({\rm{H}} \left (1,2+2\Upsilon_{1},\frac{1}{\alpha_{1}\rho}  \right )-{\rm{H}} \left (1,2+2\Upsilon_{1},\frac{2}{\alpha_{1}\rho}   \right )    \right ) \Bigg).
\end{split}
\end{aligned}\\
&\begin{aligned}
\begin{split}
\label{eq:ec_noma_w_closed}
C_{2}^{{\rm{N}}}=& -\frac{1}{\theta_{2}{n}}{\rm{ln}}\Bigg ( \epsilon+\left ( 1-\epsilon \right )   \frac{2\alpha_{1}^{-2\Upsilon_{2}}}{ \rho }e^{\psi_{2}}\Bigg ({\rm{H}} \left (1,2,\frac{2}{\rho}  \right )+\frac{n \theta_{2}\left (\alpha_{1}-1 \right)}    {\alpha_{1}{\rm{ln}2}} e^{\frac{2}{\alpha_{1}\rho}}{\rm{E_i}}\left ( -\frac{2}{\alpha_{1}\rho} \right )+ \sum_{k=2}^{\infty}  \binom{2\Upsilon_{2}}{k}\left (\frac{\alpha_{1}-1}{\alpha_{1}} \right)^k   \\  & \times \Bigg(\frac{\sum_{j=1}^{k-1} \frac{\left(j-1\right)!}{\alpha_{1}^{-j}} \left(- \frac{2}{\rho}\right)^{k-j-1}-\left(- \frac{2}{\rho} \right)^{k-1}    }{\left(k-1\right)!}             e^{\frac{2}{\alpha_{1}\rho}}    {\rm{E_i}}\left(-\frac{2}{\alpha_{1}\rho} \right)
  \Bigg)    \Bigg ) \Bigg ).
\end{split}
\end{aligned}\\
&\begin{aligned}
\begin{split}
\label{eq:ec_oma_s_closed}
{C}_{1}^{{\rm{O}}}= -\frac{1}{\theta_{1}{n}}{\rm{ln}}\Bigg ( \epsilon+\left ( 1-\epsilon \right )   \frac{2}{ \rho }e^{\frac{\psi_{1}}{2}}\left({\rm{H}} \left (1,2+\Upsilon_{1},\frac{1}{\rho}  \right )-{\rm{H}} \left (1,2+\Upsilon_{1},\frac{2}{\rho}   \right )    \right ) \Bigg ).
\end{split}
\end{aligned}\\
&\begin{aligned}
\begin{split}
\label{eq:ec_oma_w_closed}
{C}_{2}^{{\rm{O}}}= -\frac{1}{\theta_{2}{n}}{\rm{ln}}\Bigg ( \epsilon+\left ( 1-\epsilon \right )   \frac{2}{ \rho }e^{\frac{\psi_{2}}{2}}{\rm{H}} \left (1,2+\Upsilon_{2},\frac{1}{\rho}  \right ) \Bigg ).
\end{split}
\end{aligned}
\end{align}
 \hrule
\end{figure*}
\section{Numerical Results}
\label{sec:numerial_results}
We performed extensive simulations to compare the performance of the two-user NOMA and two-user OMA in finite blocklength regime. In this section, numerical results are discussed and compared, considering the users' power coefficients $\alpha_2=0.7$ and $\alpha_1=0.3$, the blocklength $n=400$, and a transmission error probability $\epsilon=10^{-6}$, unless otherwise specified.

\begin{figure*}[htb]
\minipage{0.32\textwidth}
\hspace*{0.25cm}\centering\includegraphics[width=\linewidth]{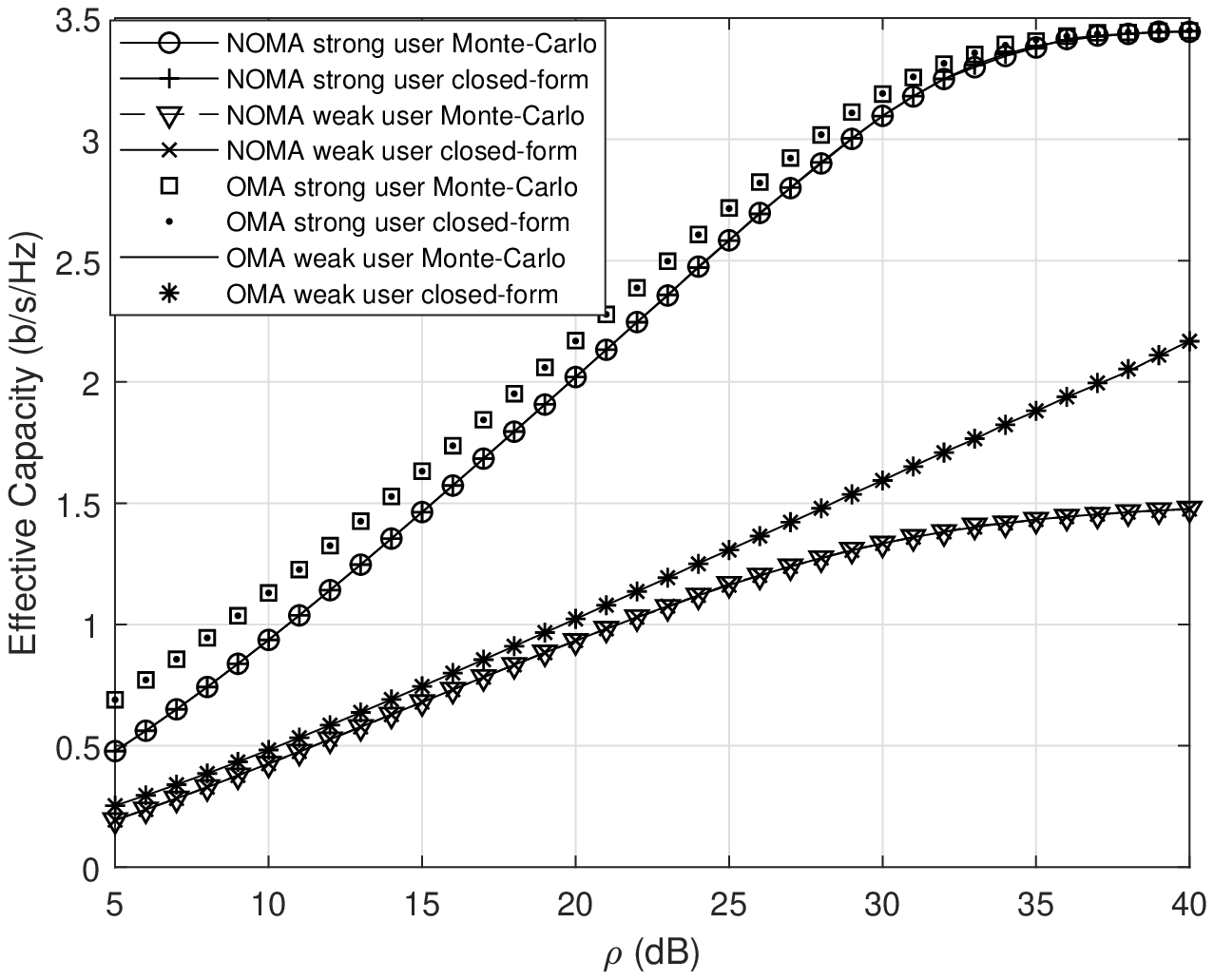}
\caption{Achievable EC of two-user NOMA and two-user OMA versus the transmit SNR.}
\label{fig1}
\endminipage\hfill
\minipage{0.32\textwidth}
\hspace*{0.25cm}\centering\includegraphics[width=\linewidth]{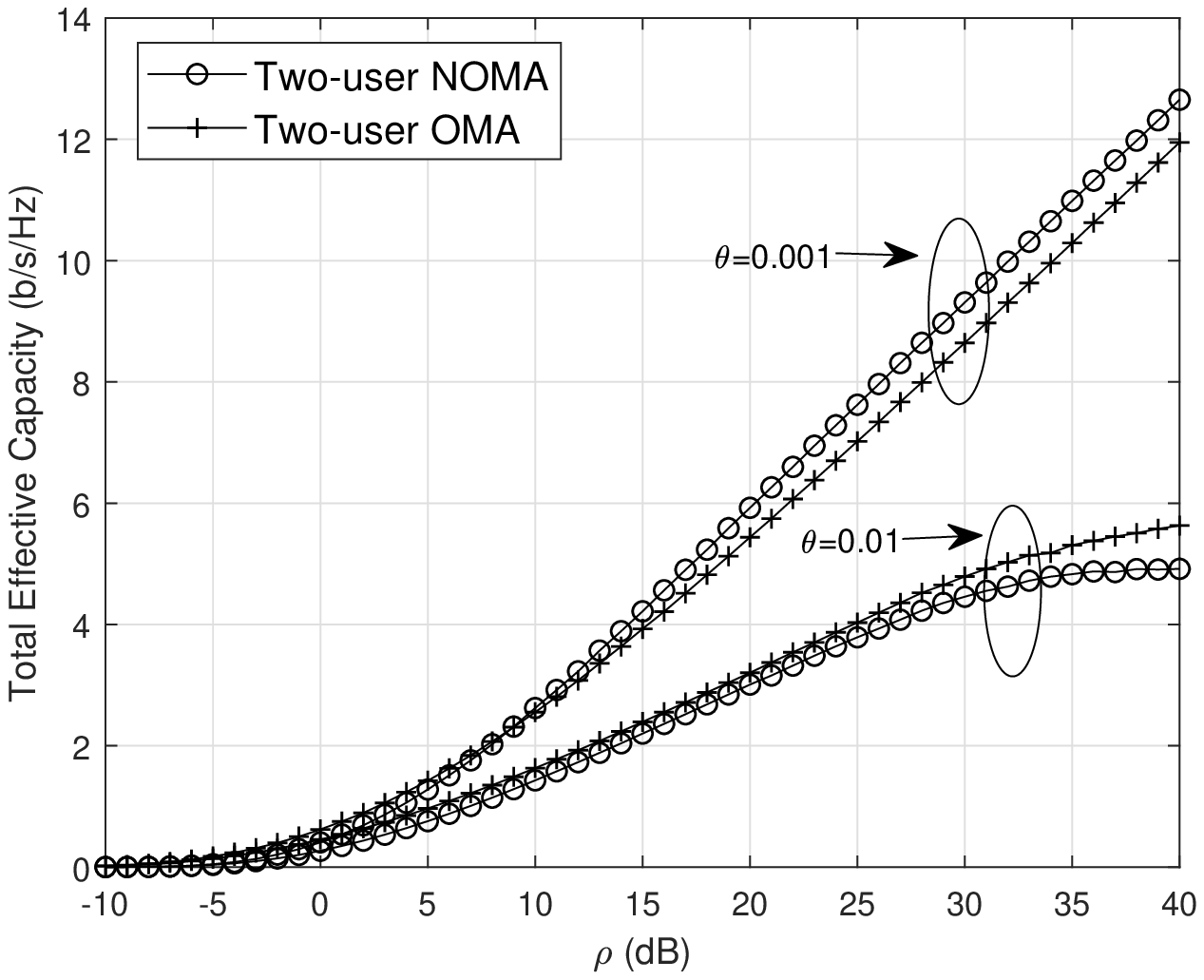}
\caption{Total achievable EC of two-user NOMA and two-user OMA versus the transmit SNR.}
\label{fig2}
\endminipage\hfill
\minipage{0.32\textwidth}%
\hspace*{0.25cm}\centering\includegraphics[width=\linewidth]{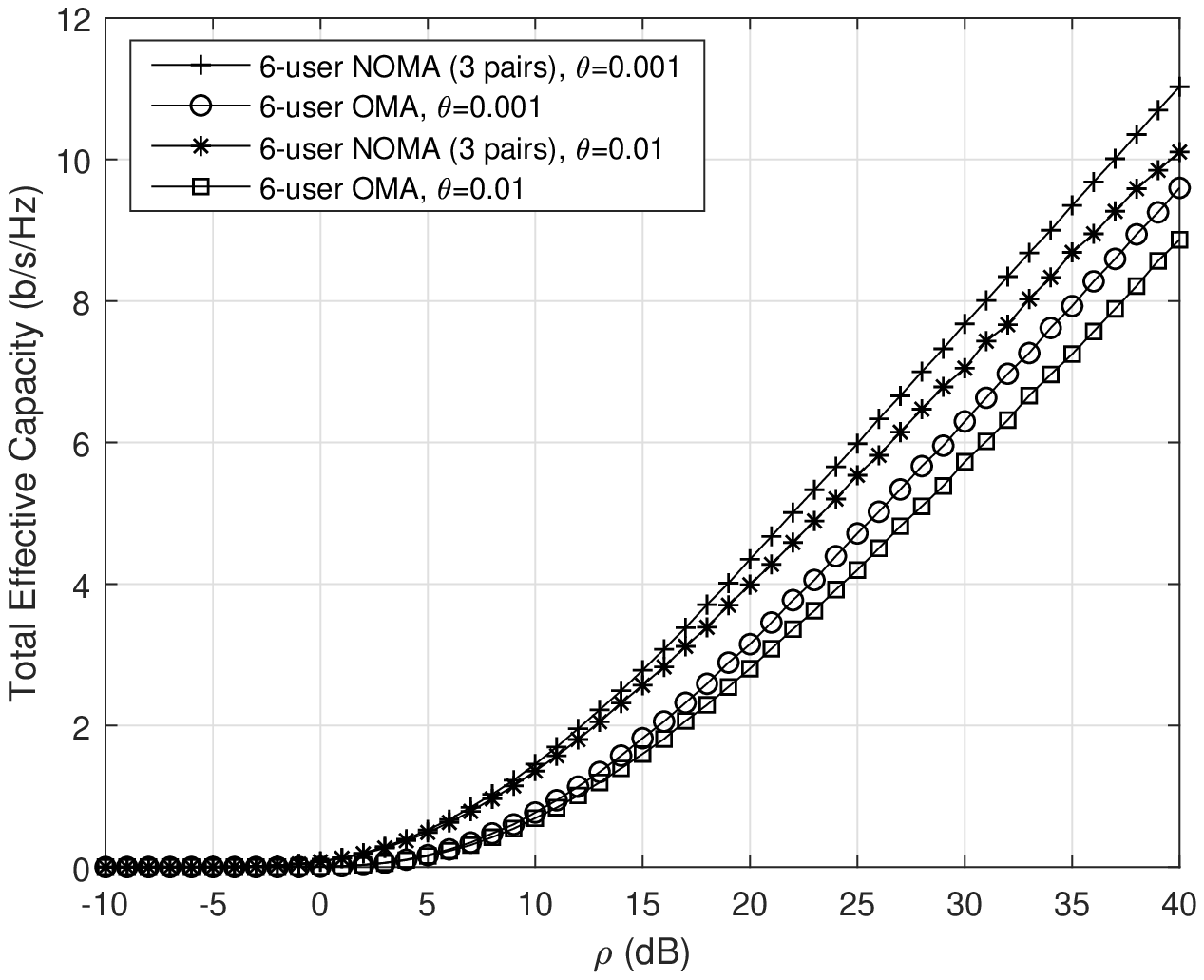}
\caption{Total achievable EC of multiple NOMA pairs and multiple OMA users versus the transmit SNR with 6 users out of 12 users. }
\label{fig3}
\endminipage
\end{figure*}

\begin{figure}[ht]
\centering
 \includegraphics[width=0.65\linewidth]{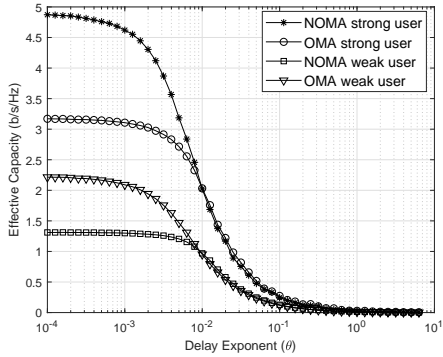}
\caption{Achievable EC of two-user NOMA and two-user OMA versus delay exponent $(\theta)$. }
\label{fig4}
\end{figure}
\begin{figure}[ht]
\centering
 \includegraphics[width=0.65\linewidth]{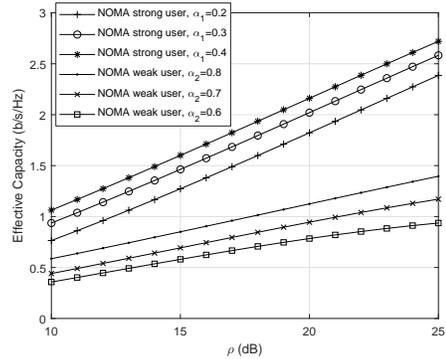}
\caption{\blue{Achievable EC of two-user NOMA versus the transmit SNR for different values of the power coefficients $(\alpha_1, \alpha_2)$.}}
\label{fig5}
\end{figure}

Fig. \ref{fig1} shows the plots of the achievable EC of two-user NOMA and two-user OMA in finite blocklength regime as a function of the transmit SNR $(\rho)$ in dB. For this evaluation, we set $\theta=0.01$. The accuracy of the derived closed-form expressions is confirmed. The figure also shows that, at very low transmit SNRs, the OMA strong user outperforms both NOMA users. However, as $\rho$ increases, the achievable EC of NOMA and OMA does not increase further and saturates at very high values of the SNR. At low SNRs, the achievable EC of the weak user is approximately the same in both NOMA and OMA, whereas at high SNRs the weak user OMA dominates with a big gap.

Fig. \ref{fig2} shows the plots of the total achievable EC versus the transmit SNR $(\rho)$.  The results reveal that the total achievable rate of NOMA outperforms the one for OMA at high SNRs when $\theta=0.001$. On the other hand, when the delay exponent becomes stringent, i.e., changes from $\theta \to 0.001$ to $\theta \to 0.01$, the total link-layer rate of OMA outperforms the one of NOMA at high SNRs. However, at the low SNRs, the total link-layer rate of NOMA and OMA are approximately the same irrespective of the delay constraints.

\blue{Fig. \ref{fig3} provides a comparative view of the total achievable EC of multiple NOMA and OMA pairs when service is provided to 6 users out of 12 possible users, and $\theta=[0.001,0.01]$. Within a pair, NOMA scheme has been implemented, while the inter-pair multiple access has been achieved using TDMA. It is clear from the curves that the multiple-user NOMA network outperforms the OMA one under different delay requirements. The figure also reveals that multiple-user NOMA and OMA perform better than the two-user access cases when the delay exponent becomes stringent.}

Fig. \ref{fig4} plots the simulation results of individual user's achievable EC of two-user NOMA and OMA versus the delay exponent $\theta$ when the transmit SNR $\rho=20$dB.  This figure shows that the NOMA users outperform the OMA users when the delay exponent is very loose. However, when the delay exponent becomes stringent, the NOMA users show a considerable loss in EC as compared to the OMA case.

\blue{Finally, Fig. \ref{fig5} shows the curves of the achievable EC of two-user NOMA versus the transmit SNR $(\rho)$ with different values of the power coefficients $(\alpha_1$,$\alpha_2)$, when $\theta=0.01$. Compared to the flexible power allocation scheme, this figure shows how the different sets of fixed power coefficients impact the performance of the two-user NOMA network.  With the increase in the power coefficients, the achievable EC of both the strong user and the weak user with NOMA also increases. This also confirms that the changing power coefficients has also a significant impact on the two-users NOMA netowrk under delay constraint.}
\section{Conclusion}
\label{sec:conclusion}
In this paper, considering NOMA and OMA with two users, we formulated the individual user's achievable EC in finite blocklength regime. We derived closed-form expressions for the individual EC of both users, in NOMA and OMA separately, and confirmed their accuracy using Monte-Carlo simulations. We investigated the performance of NOMA in comparison with OMA under heterogeneous delay QoS constraints. The performance comparison showed that at low SNRs the strong user OMA outperforms both NOMA users, while the total link-layer rate of NOMA outperforms the one of OMA at high SNRs.
\section*{APPENDIX A}
To obtain the closed-form expressions for  $C_1^{{\rm{N}}}$, ${C}_1^{{\rm{O}}}$, and $C_2^{{\rm{O}}}$, we first consider the simple case of ${C}_1^{{\rm{O}}}$. Following the order statistics from (\ref{eq:pdf_sw}), ${C}_1^{{\rm{O}}}$ (from (\ref{eq:ec_omas})) can be expanded as
\begin{align}
\label{eq:ec_oma_s_int1}
{C}_{1}^{{\rm{O}}}= -\frac{1}{\theta_{1}{n}}{\rm{ln}}\Bigg (\int_{0}^{\infty} \Big( \epsilon+ ( 1- \epsilon  )      ( &1+\gamma_1   )^{\Upsilon_{1}} e^{\frac{\psi_{1}V_{1}^O}{2}} \Big ) \nonumber \\ & \times f_{\gamma_{1:2}} ( \gamma_{1}  )   d\gamma_1   \Bigg ),
\end{align}
where $f\left ( \gamma_1 \right )=\frac{1}{\rho}e^{-\frac{\gamma_1}{\rho}}$, $F\left ( \gamma_1 \right )=1-e^{-\frac{\gamma_1}{\rho }}$, and we assume at high SNR  $\sqrt{V_{i}^{N}} \approx 1,    \sqrt{V_{i}^{O}} \approx 1$ \cite{R139}. Thus, we get
\begin{small}
\begin{align}
\label{eq:appec_oma_s_eq2}
{C}_{1}^{{\rm{O}}}= -\frac{1}{\theta_{1}{n}}{\rm{ln}}   \Bigg(\epsilon + & \left ( 1-\epsilon \right )     \frac{2}{\rho}  e^{\frac{\psi_{1}}{2} }  \Big (  \underbrace{\int_{0}^{\infty} \left( 1+\gamma_1  \right )^{\Upsilon_{1}}  e^{-\frac{\gamma_1}{\rho}}    d\gamma_s}_\text{$I_1$} \nonumber \\ &     -\underbrace{ \int_{0}^{\infty}\left ( 1+\gamma_1  \right )^{\Upsilon_{1}}  e^{-\frac{2\gamma_1}{\rho}}    d\gamma_1}_\text{$I_2$}  \Big)      \Bigg).
\end{align}
\end{small}
We recall the confluent hypergeometric function of the second kind \cite{R171}, namely,
\begin{align}
 \begin{split}
 \label{eq:app_hyper}
{\rm{H}}\left ( a,b,z \right )=\frac{1}{\Gamma\left ( a \right )} \int_{0}^{\infty}e^{-zt}t^{a-1}\left ( 1+t \right )^{b-a-1} dt.
\end{split}
\end{align}
By using (\ref{eq:app_hyper}), the integrals $I_1$ and $I_2$ shown in (\ref{eq:appec_oma_s_eq2}) can be solved as
\begin{align}
 \begin{split}
 \label{eq:app_hint12}
I_1={\rm{H}}\Big ( 1,2+\Upsilon_1,\frac{1}{\rho} \Big ), \ \ \  I_2={\rm{H}}\Big ( 1,2+\Upsilon_1,\frac{2}{\rho} \Big ).
\end{split}
\end{align}

Following the above steps, the closed-form expressions for $C_1^{{\rm{N}}}$ and ${C}_2^{{\rm{O}}}$, given in (\ref{eq:ec_noma_s_closed}) and (\ref{eq:ec_oma_w_closed}), respectively, can also be obtained.
\section*{APPENDIX B}
Following the order statistics from (\ref{eq:pdf_sw}), the achievable EC of the weak NOMA user shown in (\ref{eq:ec_nomaw}) can be expanded as
\begin{align}
\label{eq:ec_noma_w_int1}
C_{2}^{{\rm{N}}}= -\frac{1}{\theta_{2}{n}}{\rm{ln}}\Bigg (\int_{0}^{\infty}   \bigg (  \epsilon+ &( 1-\epsilon  )      \bigg ( \frac{\gamma_{2}+1 }{\alpha_1 \gamma_{2}+1 }  \bigg )^{2\Upsilon_{2}}  \nonumber  \\ &   \times   e^{\psi_{2}V_{2}^N} \bigg )   f_{\gamma_{2:2}} ( \gamma_{2} )   d\gamma_2  \Bigg ) ,
\end{align}
\noindent where $f\left ( \gamma_2 \right )=\frac{1}{\rho}e^{-\frac{\gamma_2}{\rho}}$, $F\left ( \gamma_2 \right )=1-e^{-\frac{\gamma_2}{\rho }}$,  and  we   assume at high SNR  $V_{2}^{N} \approx 1$   \cite{R139}. Thus, we get
\begin{align}
\label{eq:app_snomaeq1}
C_{2}^{{\rm{N}}}= -\frac{1}{\theta_{2}{n}}{\rm{ln}}\Bigg ( \epsilon+ ( 1-\epsilon  )   \frac{2}{\rho}e^{\psi_{2}} \int_{0}^{\infty}   & \left ( \frac{\gamma_{2}+1 }{\alpha_1 \gamma_{2}+1 }  \right )^{2\Upsilon_{2}}  \nonumber  \\ &    \times e^{-\frac{2\gamma_2}{\rho}}    d\gamma_2  \Bigg) .
\end{align}

Following the generalized binomial expansion, we can write $ \left ( \frac{\gamma_{2}+1 }{\alpha_1 \gamma_{2}+1 }  \right )^{2\Upsilon_{2}}=\left (  \frac{1} {\alpha_1} \right )^{2\Upsilon_{2}}\left (  1+ \frac{\alpha_1-1}{\alpha_1 \gamma_{2}+1 }  \right )^{2\Upsilon_{2}},$ where the expression $\left (  1+ \frac{\alpha_1-1}{\alpha_1 \gamma_{2}+1 }  \right )^{2\Upsilon_{2}}$ can further be expanded as $\left (  1+ \frac{\alpha_1+1}{\alpha_1 \gamma_{2}+1 }  \right )^{2\Upsilon_{2}}=\sum_{k=0}^{\infty}\binom{2\Upsilon_{2}}{k}\left (  \frac{\alpha_1-1}{\alpha_1 \gamma_{2}+1}\right )^{k}.$

Using the above expansions,  (\ref{eq:app_snomaeq1}) can be rewritten as
\begin{align}
\label{eq:app_snomaeq4}
 C_{2}^{{\rm{N}}}&=  -\frac{1}{\theta_{2}{n}}{\rm{ln}}\Bigg ( \epsilon+\left ( 1-\epsilon \right ) \frac{2\alpha_{1}^{-2\Upsilon_{2}}}{\rho}e^{\psi_{2}} \Bigg ( \underbrace{ \int_{0}^{\infty}       e^{-\frac{2\gamma_2}{\rho}}    d\gamma_2}_\text{$I_{a} \ (k=0) $}     \nonumber \\ &   + \underbrace{\int_{0}^{\infty}    2\Upsilon_2  \frac{\alpha_{1}-1 }{\alpha_1 \gamma_{2}+1 }  e^{-\frac{2\gamma_2}{\rho}}    d\gamma_2}_\text{$I_{b} \ ( k=1)$}  \nonumber  \\&  +     \underbrace{\int_{0}^{\infty}  \sum_{k=2}^{\infty}\binom{2\Upsilon_{2}}{k}\left (  \frac{\alpha_1-1}{\alpha_1 \gamma_{2}+1}\right )^{k}     e^{-\frac{2\gamma_2}{\rho}}    d\gamma_2}_\text{$I_c  \ (k \geq 2)$}   \Bigg ) \Bigg ).
\end{align}
Using (\ref{eq:app_hyper}), we get $I_{a}={\rm{H}} \left (1,2,\frac{2}{\rho}   \right )$. For the integrals $I_{b}$ and $I_{c}$, we use (3.353.2) and (3.352.4) from \cite{R172}, which yields
\begin{align}
\label{eq:app_t7}
\int_{0}^{\infty}\frac{e^{-zt}}{t+b}d_{t}=-e^{bz}{\rm{E_i}}(-bz),  \left [\left | \rm{arg} \ b \right |<\pi, \ \rm{Re}(z)>0 \right ],
\end{align}
\noindent and
\begin{align}
\begin{split}
\label{eq:app_t8}
&\int_{0}^{\infty}\frac{e^{-zt}}{(t+b)^l}d_{t}=\frac{1}{(l-1)!}\sum_{j=1}^{l-1} \left ( j-1 \right )!(-z)^{l-j-1} (b^{-j}) \\ &
-\frac{(-z)^{l-1}}{(l-1)!}e^{bz}{\rm{E_i}}(-bz), \ \  \left [l\geq2, \ \left  | \rm{arg} \ b \right |<\pi, \ \rm{Re} \ z >0 \right ],
\end{split}
\end{align}
\noindent where ${\rm{E_i}}(.)$ is the exponential integral \cite{R172}. Finally, using (\ref{eq:app_t7}) and (\ref{eq:app_t8}), the closed-form expression for $C_2^{{\rm{N}}}$ can be found as shown in (\ref{eq:ec_noma_w_closed}).
\bibliographystyle{IEEEtran}

\begin{thebibliography}{10}

\bibitem{R173}
W.~Saad, M.~Bennis, and M.~Chen, ``A vision of {6G} wireless systems:
  Applications, trends, technologies, and open research problems,'' \emph{arXiv
  preprint arXiv:1902.10265}, July 2019.

\bibitem{R160}
M.~{Amjad} and L.~{Musavian}, ``Performance analysis of {NOMA} for
  ultra-reliable and low-latency communications,'' in \emph{IEEE Globecom
  Workshops (GC Wkshps)}, Abu Dhabi, Dec. 2018, pp. 1--5.

\bibitem{R166}
X.~{Sun}, S.~{Yan}, N.~{Yang}, Z.~{Ding}, C.~{Shen}, and Z.~{Zhong},
  ``Short-packet downlink transmission with non-orthogonal multiple access,''
  \emph{IEEE Trans. Wireless Commun.}, vol.~17, no.~7, pp. 4550--4564, Apr.
  2018.

\bibitem{R177}
M.~Amjad, L.~Musavian, and S.~A{\"\i}ssa, ``Effective capacity of {NOMA} with
  finite blocklength for low-latency communications,'' \emph{arXiv preprint
  arXiv:2002.07098}, Feb. 2020.

\bibitem{R139}
Y.~Polyanskiy, H.~V. Poor, and S.~Verd{\'u}, ``Channel coding rate in the
  finite blocklength regime,'' \emph{IEEE Trans. Inf. Theory}, vol.~56, no.~5,
  pp. 2307--2359, May 2010.

\bibitem{R140}
Y.~Yu, H.~Chen, Y.~Li, Z.~Ding, and B.~Vucetic, ``On the performance of
  non-orthogonal multiple access in short-packet communications,'' \emph{IEEE
  Commun. Lett.}, vol.~22, no.~3, pp. 590--593, Mar. 2018.

\bibitem{R176}
E.~Dosti, M.~Shehab, H.~Alves, and M.~Latva-aho, ``On the performance of
  non-orthogonal multiple access in the finite blocklength regime,'' \emph{Ad
  Hoc Netw.}, vol.~84, pp. 148--157, Mar. 2019.

\bibitem{R179}
M.~Bello, ``Asymptotic regime analysis of {NOMA} uplink networks under {QoS}
  delay constraints,'' \emph{arXiv preprint arXiv:2001.11423}, Jan 2020.

\bibitem{R156}
M.~Shehab, H.~Alves, and M.~Latva-aho, ``Effective capacity and power
  allocation for machine-type communication,'' \emph{IEEE Trans. Veh.
  Technol.}, vol.~68, no.~4, pp. 4098--4102, Apr. 2019.

\bibitem{R178}
M.~Shehab, E.~Dosti, H.~Alves, and M.~Latva-aho, ``On the effective capacity of
  {MTC} networks in the finite blocklength regime,'' in \emph{IEEE European
  Conf. on Networks and Commun. (EuCNC)}, Jun. 2017, pp. 1--5.

\bibitem{R175}
\emph{Study on Downlink Multiuser Superposition Transmission for {LTE},
  {3GPP}}, Shanghai, China,, Mar. 2015.

\bibitem{R163}
H.~A. David and H.~N. Nagaraja, ``Order statistics,'' \emph{Encyclopedia of
  Statistical Sciences}, 2004.

\bibitem{R171}
M.~Abramowitz and I.~A. Stegun, ``Handbook of mathematical functions dover
  publications,'' \emph{New York}, p. 361, 1965.

\bibitem{R172}
I.~S. Gradshteyn and I.~M. Ryzhik, \emph{Table of integrals, series, and
  products}.\hskip 1em plus 0.5em minus 0.4em\relax Academic press, 2014.

\bibitem{R158}
M.~{Amjad}, L.~{Musavian}, and M.~H. {Rehmani}, ``Effective capacity in
  wireless networks: A comprehensive survey,'' \emph{IEEE Commun. Surveys
  Tuts.}, Jul. 2019.

\bibitem{R167}
C.-S. Chang, \emph{Performance guarantees in communication networks}.\hskip 1em
  plus 0.5em minus 0.4em\relax Springer Science \& Business Media, 2012.

\bibitem{R154}
M.~C. Gursoy, ``Throughput analysis of buffer-constrained wireless systems in
  the finite blocklength regime,'' \emph{EURASIP J. Wireless Commun. Net.},
  vol. 2013, no.~1, p. 290, Dec. 2013.

\end{thebibliography}

\end{document}